\documentclass{article}

\usepackage[final, nonatbib]{wdcs2020} 

\usepackage[utf8]{inputenc} 
\usepackage[T1]{fontenc}    
\usepackage{hyperref}       
\usepackage{booktabs}       
\usepackage{amsfonts}       
\usepackage{nicefrac}       
\usepackage{microtype}      
\usepackage{changepage}
\usepackage{makecell}
\usepackage{placeins}
\usepackage{xcolor}
\PassOptionsToPackage{hyphens}{url}\usepackage{hyperref}
\usepackage{booktabs,caption}
\usepackage[flushleft]{threeparttable}
\usepackage{url}

\title{Ethical Testing in the Real World: Evaluating Physical Testing of Adversarial Machine Learning}

\author{%
Kendra Albert\thanks{All authors contributed equally. The list of authors is arranged alphabetically} \\
Harvard Law School\\
  \texttt{kalbert@law.harvard.edu}
\And
  Maggie Delano\footnotemark[1]\\
Swarthmore College\\
  \texttt{mdelano1@swarthmore.edu}
\And
  Jonathon Penney\footnotemark[1]\\
 Citizen Lab\\
  \texttt{jon@citizenlab.ca} \\
  \And
 Afsaneh Rigot\footnotemark[1]\\
Article 19\\
  \texttt{arigot@cyber.harvard.edu} \\
  \And
Ram Shankar Siva Kumar\footnotemark[1]\\
Microsoft\\
  \texttt{ramk@microsoft.com} \\
}

\begin{document}

\maketitle

\begin{abstract}
This paper critically assesses the adequacy and representativeness of physical domain testing for various adversarial machine learning (ML) attacks against computer vision systems involving human subjects. Many papers that deploy such attacks characterize themselves as ``real world.'' Despite this framing, however, we found the physical or real-world testing conducted was minimal, provided few details about testing subjects and was often conducted as an afterthought or demonstration. Adversarial ML research without representative trials or testing is an ethical, scientific, and health/safety issue that can cause real harms. We introduce the problem and our methodology, and then critique the physical domain testing methodologies  employed by papers in the field. We then explore various barriers to more inclusive physical testing in adversarial ML and offer recommendations to improve such testing notwithstanding these challenges.
\end{abstract}

\section{Introduction}
Advances in machine learning (ML) techniques have enabled computer vision systems such as facial and person recognition to erode expectations of privacy and anonymity online and in person. These systems are already being employed to track dissidents and identify protesters \cite{garvie2019america, mozur2019hong}. Adversarial ML attacks are one way for individuals to combat surveillance. Unfortunately, the limited adversarial ML attacks that exist are not yet ready for the ``real world'' \cite{hill2020}. In this paper, we turn a critical lens on the adequacy and representativeness of physical domain testing for various adversarial ML attacks against computer vision systems involving human subjects. Many papers that use physical adversarial attacks characterize themselves as ``real world''or ``practical'' \cite{komkov2019advhat, wu2019making, zhou2018invisible} and many receive glowing write ups in popular news outlets (see Table \ref{tab:paperSummary2}). In practice, however, we found that physical or real-world testing was minimal, papers provided few details about testing subjects, and testing was often conducted as an afterthought or as a demonstration.  The hype around adversarial ML attacks have tangible implications: multiple authors of this paper have been approached by advocates who wish to use adversarial machine learning techniques in high risk, real world contexts.

Representative testing is an ethical, scientific, and health/safety issue. Adversarial research without representative trials or ethically limited testing violate the central moral principles at the heart of research ethics—autonomy, beneficence, non-maleficence, and justice— \cite{beauchampPrinciplesBiomedicalEthics2001,BelmontReport1979}—and can cause real harms, not only to participants. Those harms are not just unreliable and unscientific results or vulnerable and less secure ML systems, but that failing to meaningfully test these technologies may result in the publication of results that contribute to broader societal problems of systemic racism, discrimination, prosecution, and marginalization, particularly for vulnerable minority groups and communities. 

We first describe our methodology, including the criteria used to select three different categories of adversarial ML papers for review, and then critique the physical domain testing methodologies they employed. Having set out those critiques, we explore various barriers to more diverse and representative study populations and testing in adversarial ML, and offer recommendations to improve such testing in notwithstanding these challenges.

\begin{table}

\caption{List of Papers Analyzed}
    \centering
    \begin{tabular}{c c c c c}
    \toprule
        Paper & Type & Sample Size & Demographics \\ \midrule
        \cite{thys2019fooling} & Adversarial Patch & Unknown & None \\ 
        \cite{zhou2018invisible} & Adversarial Cap & 1 & None  \\ 
        \cite{komkov2019advhat} & Adversarial Cap & 10 & Age, Gender \\  
       \cite{Xu2019AdversarialTE}  & Adversarial T-Shirt & 5 & Gender \\  
        \cite{wu2019making} & Adversarial T-Shirt & 1 & None \\  
        \cite{sharif2019general} & Adversarial Eyewear & 3 & Age, Gender, Race\\ 
        \cite{chen2017targeted} & Adversarial Eyewear & 5 & None \\
    \bottomrule
    \end{tabular}
    \label{tab:paperSummary}
\end{table}

\section{Physical Adversarial Attack Study Methods}

In our evaluation, we focused on authors who chose to demonstrate physical adversarial attacks in the computer vision domain involving human subjects, and excluded non-human subjects like stop signs \cite{eykholtRobustPhysicalWorldAttacks2017}. For a full list of papers analyzed, see \textbf{Table 1}. We paid particular attention to papers that got widespread press coverage (see\textbf{ Appendix}), independent of their publication status. This resulted in three different categories of papers: adversarial hats, adversarial clothing/patches, and adversarial eyewear -- each of which when worn, could evade person detection in general, or facial recognition, in particular. 

To be clear, we do not evaluate the veracity of the results in the digital domain or on datasets.  We merely attempt to understand the testing protocol followed when testing the attacks in the physical domain, and point out how such testing falls short of what might be required to make generalizable claims. Below, we summarize a number of key critiques of the physical testing of adversarial attacks.

\subsection{Critique 1: Digital != Physical}
Scholarship around physical adversarial ML generally evaluate the efficacy of their attacks in two steps: results are demonstrated against a digital dataset (which in our paper we call a ``digital attack'') as well as results are shown in the actual physical domain (which in our paper we call a ``physical attack''), often accompanied by a photograph.  

Perhaps the most important point with the human testing physical adversarial attack literature is that testing in the digital domain or using a wide variation of digital datasets does not eliminate the need for testing in the physical domain. In fact, some of the papers themselves make this point quite directly, such as Wu et al. and Xu et al., who both noted the divergence between physical attacks and digital attacks \cite{wu2019making,Xu2019AdversarialTE}.  

But that message has not universally sunk in. In some papers we reviewed, the authors did not robustly test the tech in real life, or if they did, did not describe it in a way that allowed for either evaluation or replication of their methods. For example, in Thys et al., the authors created patches to attack person detection \cite{thys2019fooling}. Their primary testing was on the Inria dataset, and they digitally transformed the patches and inserted them into existing images. They  then did a real world test. It is unclear how that test was controlled, who it was on, or whether there was any IRB approval for the physical testing. Nonetheless,  the authors made broad claims regarding the real applications of their work: ``From our real-world test with printed out patches we can also see that our patches work quite well in hiding persons from object detectors, suggesting that security systems using similar detectors might be vulnerable to this kind of attack'' \cite{thys2019fooling}. 

Other papers, such as Wu et al, do a better job of explaining their physical methodology or cabining their analysis \cite{wu2019making}. But even so, their public claims seem to suggest broad applicability of their technology that is not based on what their testing demonstrates \cite{RageOnWorldLargest}. Judiciousness in non-paper claims is vital, because in real world contexts where people may deploy these adversarial ML attacks or tools to evade detection or protect themselves, they may face great risks and danger if those attacks are not reliably or adequately tested.

\subsection{Critique 2: No Consent Procedures and Potential Harm to Participants}

None of the studies we reviewed discussed consent procedures or formal approval. This is striking because when human subjects are involved in research, United States researchers are legally required to consult an institutional review board (IRB). Although digital research with human subjects is also subject to IRB approval, physical testing is the reason why IRBs were created.

Without formal consent procedures, it may have been unclear to test subjects what the authors would do with their images or with the results, and the options for withdrawal of consent may not have been articulated. In addition, it may have been unclear what harms might have resulted from participation in the various studies.

Only one of the papers we studied discussed potential harm to participants. This paper involved pointing IR-producing diodes at someone’s face, and noted that the authors ``didn't recruit a large number of volunteers to participate in our experiments, because we can not afford potential medical risks resulted from IR'' \cite{zhou2018invisible}.

As above, it is possible that the lack of IRB approval and/or discussion of harm to participants was because the authors did not view the users of these technologies as study participants, either because they tested upon themselves or on friends. This may also explain the varying approaches to blurring faces - some papers, such as the one involving the Adversarial T-Shirt in \cite{Xu2019AdversarialTE}, blurred the participant’s images, but most did not. However, if authors are making generalized claims as to the use of their technologies in the real world, it is unethical to fail to consult an IRB.

\subsection{Critique 3: Small Sample Size, Low Sample Diversity, and Other Sampling Problems}

Finally, all of the papers we reviewed had noticeably small sample sizes. Even when there is a wide variation in testing environments, the studies only included data from one or two people \cite{wu2019making, zhou2018invisible}. With many of these technologies, the characteristics of the test population, including their gender, race, and age could inform the generalizability of the results. Many papers do not even report this information (see Table \ref{tab:paperSummary}). 

For papers where it was unclear who the test population was, we reached out to the authors to get more information and we found a significant lack of participant diversity. That is, because authors were often testing on themselves or friends/colleagues, the sample was one of convenience rather than a deliberate choice.

Across many papers, there was no opportunity to observe whether there even could be variance across human test subjects. For some technologies, we could theorize a physical way in which an attack might interact with different skin colors, skin textures, etc. Likewise, in clothing testing, there was rarely variation in the person’s weight and height, or how the clothes were worn, meaning that it is unclear how widely generalizable the results are to other circumstances \cite{Xu2019AdversarialTE}. 

In our view, this is a significant flaw in the approach of physical attack testing or demonstrations - as readers, we do not even know whether the characteristics of the test population matter. Individuals historically excluded from research studies may have similar or divergent experiences from those historically included \cite{mosesonImperativeTransgenderGender2020, tannenbaumSexGenderAnalysis2019}.

\section{Barriers to More Representative Testing and Recommendations}

Adversarial ML research without representative trials or testing that is then relied upon by other researchers, industry, government, and the press, can cause real harms. In this section, we will discuss some potential  barriers to improved testing, along with recommendations for addressing these barriers and creating more representative trials for adversarial ML research. Some of these recommendations are specific to adversarial ML, but others address broader harms and challenges resulting from non-representative trials and testing.  With them, we hope to open a dialogue around ways adversarial ML research can be conducted with further diversity, value alignment, and harm awareness. 

\subsection{Lack of Diversity}

 Many of the research participants were the authors themselves and/or colleagues. If participants in adversarial ML research are primarily also adversarial ML researchers themselves, a systemic barrier to more representative trials and testing in adversarial ML research is the lack of diversity in the research community itself  \cite{nationalsciencefoundationIndicators2018,crawfordAINowReport2019}. These diversity challenges may be even greater within adversarial ML because adversarial ML remains a new and ``emerging'' ML field wherein expertise and training is arguably even more specialized and discrete than the broader ML field \cite{rahimiAdversarialMachineLearning2020}.

\textbf{Recommendation to Researchers}: 
Lack of diversity in adversarial ML research community should not stop diversifying the test pool based on gender, racial and cultural diversity. More directly, it is especially important to include participants disproportionately subject to surveillance such as Black people and other people of color, trans, non-binary and gender nonconforming people, sex workers, and those at the intersections of these identities. Testing/trials should also include anthropometric factors (height, weight, body shape and size), and ability (mobility, vision, etc.) relevant to their research project, and test in varied real world environments with different perspectives (front, side, back, etc.). Contextualize the research in terms of the population who might seek to use it; different populations will have different cultural contexts \cite{article19AppsArrestsAbuse2018}. In addition to including a wide group  of participants in research studies (no more testing on 1 participant), it is essential to carefully document methods used to enable replication by other researchers.

\subsection{Lack of Infrastructure / Support / Incentives}
Lack of infrastructure, support, and incentives for increasing diversity in research and testing are all are common barriers in fields including and beyond ML \cite{hamelBarriersClinicalTrial2016}. One challenge in the adversarial ML field is that most research escapes ethical oversight by traditional research infrastructure like ethical review committees (ERCs) and institutional review boards \cite{mcquillanPeopleCouncilsEthical2018}. These research ethics bodies in universities should, in the normal course, raise issues related to consent in research testing and design. However, algorithmic methods, like those employed in adversarial attacks, often escape oversight when data sets are public, pre-existing, or ``the abstraction inherent in machine learning'' obscures links to human subjects or \cite{mcquillanPeopleCouncilsEthical2018,metcalfWhereAreHuman2016}. Without human subjects, IRB oversight can be circumvented.

Even where IRBs do provide oversight, they have failed to address some diversity issues, in part due to the same diversity challenges in wider academia, but also other systemic problems as well—like lack of expertise and training among IRB and ERC members \cite{lencaConsiderationsEthicsReview2018}. In ``big data'' research like adversarial ML, privacy and data-protection issues tend to dominate IRB and ERC concerns, with little priority given to issues of fairness and risks of discrimination \cite{lencaConsiderationsEthicsReview2018}. Industry lacks even these most basic research and ethical infrastructure and the few efforts to establish similar such institutions, like Facebook’s internal IRB or Google’s Advisory Council, have failed \cite{crawfordAINowReport2019}.

Finally, there are often no incentives across corporate, academic, and governmental sectors for stronger diversity, representative study and testing groups in research. While the NSF, ACM\cite{noauthor_its_2018} (and even NeurIPS), requires Broader Impacts statements, these statements rarely reflect deep consideration of the consequences of the research. Additionally, research with negative or non-significant findings--like an unsuccessful attack-- may not be publishable. Works neglecting these issues are also rarely penalized: with growing expectations for the amount and quality of publications for advancement, academics often have little incentive to conduct studies with larger, more diverse populations, rather than moving on to the next paper.

\textbf{Recommendation to Institutions and Research Journals}  Publication venues should place stringent requirements on researchers to clearly highlight the limitations of their trials in the methodology, introduction and conclusions of papers. For example, PETS, a security conference, requires authors acknowledge the limitations of their work with regards to particular populations \cite{PETSConference2020}.

\section{Conclusion}
"Invisibility cloak". "Invisible Mask" \cite{zhou2018invisible, wu2019making}. Glowing media reports for adversarial attacks on ML systems designed to evade facial recognition technology or object/person detectors abound. For the ``Invisibility Cloak,'' the New Yorker published a lengthy story on point entitled ``Dressing for the Surveillance Age'' \cite{seabrookDressingSurveillanceAge} and Ars Technica declared, ``Some shirts hide you from cameras—but will anyone wear them?'' \cite{ShirtsHideYou}. The ``invisible mask'' was heralded by Boing Boing as a tool that ``can fool facial recognition software into thinking anyone is anyone else'' \cite{doctorowInvisibleTargetedInfrared2018}. Yet each was only tested in the physical domain on one person, raising serious questions about their reliability in real world contexts.

We acknowledge that the media frequently exaggerates ML research, and though the researchers have titled their paper such, no where do the researchers claim they invented an invisibility cloak or mask. When physical adversarial ML research garners a lot of attention, researchers must be deliberate in their methodology. Computer scientists and ML researchers, trained in dataset-driven testing, may not have the skills or desire to do extensive human subjects research as part of their work, let alone conduct representative, diverse trials. However, if adversarial ML researchers wish to make claims about ``real world'' attacks, \textit{actual} real world testing is necessary.

\section*{Broader Impact}



Our paper is proposing increased physical testing of adversarial ML techniques, especially with participants already disproportionately subject to surveillance such as Black people and other people of color, trans, non-binary and gender nonconforming people, sex workers, and those at the intersections of these identities. Increased physical testing of adversarial ML techniques, especially with groups from diverse backgrounds, will increase knowledge about the effectiveness of these techniques across populations, potentially leading to improved understanding and effectiveness of adversarial ML attacks.

Our suggestions, if followed, will produce more robust and representative physical testing of adversarial ML attacks involving human subjects. This, in turn, may produce more effective adversarial ML attacks that can help protect those historically marginalized from invasive surveillance, especially through facial recognition technologies. At the very least, we hope that our paper will result in increased discussion of the limitations of current physical testing of adversarial ML attacks, which should improve the field as a whole. Our paper suggests changes in how the adversarial ML field approaches its work to better distribute its benefits and produce more rigorous science.

If our recommendations are taken, and adversarial ML researchers improve their testing, any increased effectiveness of adversarial ML attacks will ``work both ways'' -- the same techniques developed to prevent surveillance could be used against these same groups. Adversarial ML attacks could also be used by those evading law enforcement. Overall, we question the need for state surveillance of the magnitude experienced today, and consider more agency among individual people to outweigh state security concerns. 

Another potential concern is that researchers who lack experience with physical testing with marginalized populations may cause harm while attempting to include members of those populations in their testing. However, we hope that IRBs or ERCs may provide a stop gap for this, as well as to address this in future work. 

\begin{ack}
We would like to thank Bruce Schneier for his continued support and encouragement of this team and our work. This work did not receive specific financial support.
\end{ack}

\raggedright
\bibliographystyle{plain}

\bibliography{citations}

\pagebreak

\section*{Appendix} 

\begin{table}[h]
\caption{Media Coverage}
    \centering
    \begin{tabular}{c c c c}
    \toprule
        Paper & Type & Sample Size & Media Coverage \\ \midrule
        \cite{thys2019fooling} & Adversarial Patch & Unknown & Mashable \cite{dermentzi_ai_nodate}, Verge \cite{vincent_this_2019}, Business Insider \cite{holmes_these_nodate}  \\ 
        \cite{zhou2018invisible} & Adversarial Cap & 1 & Boing Boing \cite{doctorowInvisibleTargetedInfrared2018}, DigitalTrends \cite{dormehl_led_2018} \\ 
        \cite{komkov2019advhat} & Adversarial Cap & 10 & Synced \cite{noauthor_adversarial_nodate-1} \\  
       \cite{Xu2019AdversarialTE}  & Adversarial T-Shirt & 5 & VentureBeat \cite{wiggers_researchers_2019} \\  
        \cite{wu2019making} & Adversarial T-Shirt & 1 & NewYorker \cite{seabrookDressingSurveillanceAge}, Mashable \cite{month_future_2020}, ArsTechnica \cite{ShirtsHideYou} \\  
        \cite{sharif2019general} & Adversarial Eyewear & 3 & Quartz \cite{gershgorn_all_nodate}, Vice \cite{margolin_these_nodate} \\ 
        \cite{chen2017targeted} & Adversarial Eyewear & 5 & NPR \cite{noauthor_adversarial_nodate} \\ 
    \bottomrule
    \end{tabular}
    \label{tab:paperSummary2}
\end{table}

\end{document}